\documentclass[sigconf,nonacm,natbib=true,pbalance]{acmart}
\AtBeginDocument{%
  }

\usepackage{csquotes} 
\usepackage{pbalance}
\usepackage{mdframed}
\usepackage{enumitem}
\usepackage{tabularx}   

\usepackage{balance}
\usepackage{booktabs}   
\usepackage{makecell}   
\renewcommand{\arraystretch}{1.2}  

\usepackage{xcolor}
\usepackage{caption}
\usepackage{multirow}
\usepackage{listings}

\newcommand{\code}[1]{{\small\texttt{#1}}}

\newmdenv[
  backgroundcolor=gray!8,
  linecolor=gray!50,
  roundcorner=4pt,
  skipabove=6pt,
  skipbelow=6pt,
  innerleftmargin=8pt,
  innerrightmargin=8pt,
  innertopmargin=6pt,
  innerbottommargin=6pt
]{prompt}

\makeatletter
\newcommand{\acmrightssize}{\fontsize{8}{9.5}\selectfont}

\newcommand{\firstpagerights}[1]{%
  \begingroup
    \renewcommand\thefootnote{}%
    \footnotetext{%
      \acmrightssize
      \raggedright
      \setlength{\parskip}{0pt}%
      \setlength{\parindent}{0pt}%
      #1%
    }%
    \addtocounter{footnote}{0}%
  \endgroup
}
\makeatother

\begin{document}

\title[UXSim: Towards a Hybrid User Search Simulation]{UXSim: Towards a Hybrid User Search Simulation}

\author{Saber Zerhoudi}
\orcid{0000-0003-2259-0462}
\affiliation{%
  \institution{University of Passau}
  \city{Passau}
  \country{Germany}
}
\email{saber.zerhoudi@uni-passau.de}

\author{Michael Granitzer}
\orcid{0000-0003-3566-5507}
\affiliation{%
  \institution{University of Passau}
  \city{Passau}
  \country{Germany}
}
\affiliation{%
  \institution{IT:U Austria}
  \city{Linz}
  \country{Austria}
}
\email{michael.granitzer@uni-passau.de}

\renewcommand{\shortauthors}{S. Zerhoudi et al.}

\begin{abstract}
Simulating nuanced user experiences within complex interactive search systems poses distinct challenge for traditional methodologies, which often rely on static user proxies or, more recently, on standalone large language model (LLM) agents that may lack deep, verifiable grounding. The true dynamism and personalization inherent in human-computer interaction demand a more integrated approach. This work introduces UXSim~\footnote{\url{https://searchsim.org/uxsim}}, a novel framework that integrates both approaches. It leverages grounded data from traditional simulators to inform and constrain the reasoning of an adaptive LLM agent. This synthesis enables more accurate and dynamic simulations of user behavior while also providing a pathway for the explainable validation of the underlying cognitive processes.
\end{abstract}



\begin{teaserfigure}
  \begin{center}
\includegraphics[width=\textwidth]{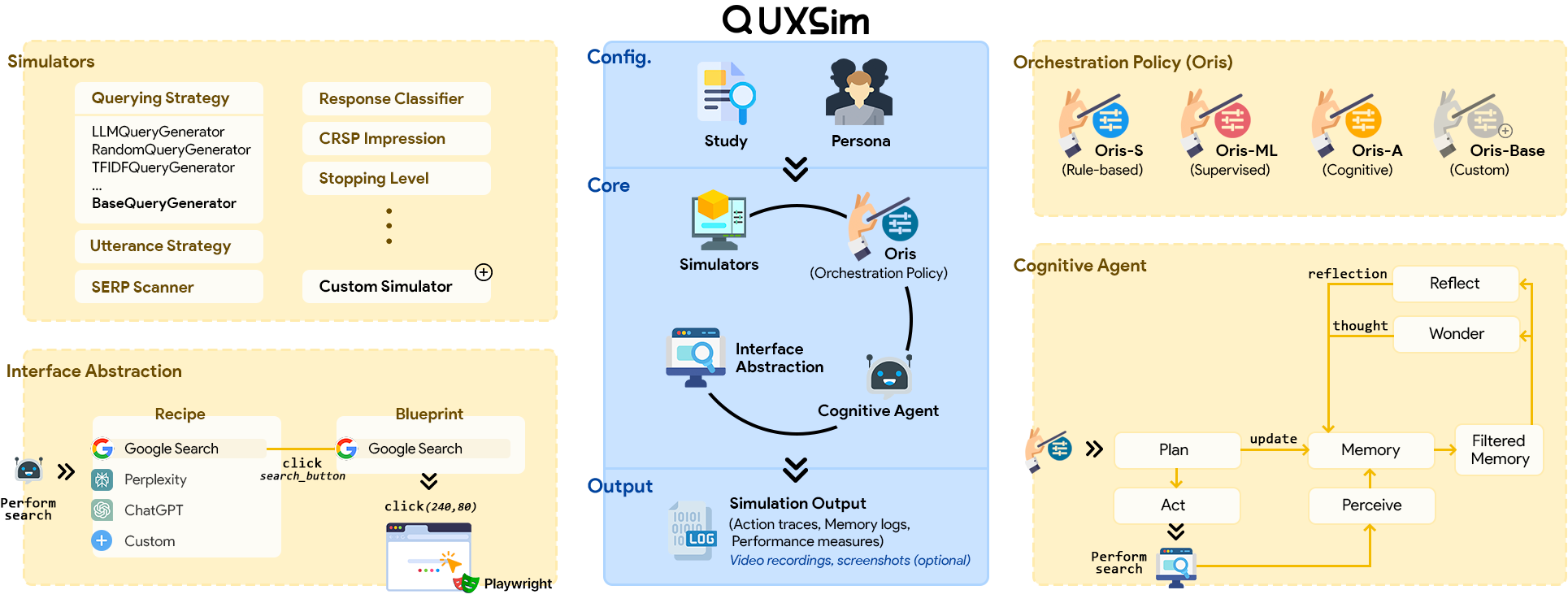}
\vspace{-3mm}
  \caption{Architecture of the UXSim framework. The Oris policy orchestrates the simulation by invoking either Simulators or capabilities within the Cognitive Agent, while the Interface Abstraction layer grounds the resulting action on the UI.}
  \label{fig_uxsim}
\end{center}
\end{teaserfigure}

\maketitle
\enlargethispage{2\baselineskip}
\firstpagerights{%
  © ACM, 2025. This is the author's version of the work.\\
  The definitive version was published in:
  \emph{Proceedings of the 34th ACM International Conference on Information and Knowledge Management\\
  (CIKM'25), November 10--14, 2025, Seoul, Republic of Korea}.\\
  DOI: \url{https://doi.org/10.1145/3746252.3761640}
}

\section{Introduction}
User simulation is a key method in Information Retrieval (IR) for evaluating systems at scale. A major challenge is moving beyond simulating simple actions, like clicks and queries, to modeling the complex thought processes that drive them. To create truly realistic user models, a simulator must capture the hidden reasoning and decision-making behind a user's observable behavior.

Research has generally followed two paths, each with distinct drawbacks. Traditional, rule-based simulators are structured and predictable but are too simple to capture dynamic human thinking. In contrast, modern Large Language Models (LLMs) can reason like humans but often produce unrealistic or ungrounded behavior when used as standalone agents, as they tend to pursue idealized solutions rather than exhibiting plausible human behavior.

This paper introduces UXSim, an open-source framework designed to solve this problem by combining the two approaches. It provides a structure where the grounded, predictable outputs from traditional simulation components are used to guide the flexible reasoning of an LLM. This synthesis results in user behavior that is both adaptive and plausibly connected to the task. UXSim is designed as a modular, \textbf{U}ser-friendly, and e\textbf{X}tensible tool for \textbf{Sim}ulation. It is supported by a web interface that lowers the barrier to entry and simplifies experimentation, analysis, and the ability to share simulation runs without exposing sensitive training data.

\section{Related Works}
Foundational toolkits like SimIIR~\cite{MaxwellA16, Zerhoudi0PBSHG22, Azzopardi00KM0P24} established a component-based approach to simulating full search sessions, including query generation, result examination, and stopping decisions. This modular design remains common today. The recent rise of LLMs has led to more powerful simulators like USimAgent~\cite{ZhangWGLM24} and BASES~\cite{RenQQLZ0WW24}, which use language models to generate more complex and nuanced user behaviors. Specialized frameworks have also emerged for areas like conversational search (e.g., UserSimCRS~\cite{AfzaliDB023}) and recommender systems (e.g., RecSim~\cite{Eugene19}). These diverse tools typically share a common modular design, with separate components for query generation, click modeling, and stopping decisions.

Early simulation work focused on probabilistic models like the User Browse Model (UBM)~\cite{DupretP08} and cognitive architectures like ACT-R~\cite{ritter2019act}. Query reformulation has also been modeled explicitly; for example, query change as a contextual Markov model for simulating user search behaviour~\cite{Zerhoudi:2021:FIRE}. These models were important for establishing the field but relied on simplified rules that could not fully capture fluid human thought. A related challenge lies in the simulated environments themselves. Past studies often used simplified interfaces that lacked the complex visual elements—like advertisements, knowledge panels, or interactive filters—that are known to influence user behavior on modern search pages. Interacting directly with such real-world UIs is a challenge for standard automation tools, which are often brittle. UXSim is designed to solve this problem with a stable, semantic abstraction layer that provides a robust and efficient connection for simulations on complex, modern interfaces.

More recent work has applied LLMs directly as user agents~\cite{WangZYCTZCLSSZXDWW25, Zhang25}. While these agents can generate human-like ``thoughts'' to rationalize their actions, their reasoning can be disconnected from the task context. They are generally not designed to incorporate the structured constraints that traditional simulators handle well, such as user personas or behavioral histories. UXSim is designed specifically to fill this gap, creating a research framework where the generative power of LLMs is purposefully grounded by data and context from other simulation components.

\section{The UXSim Framework}
This section presents the UXSim framework from three perspectives. We first outline its high-level architecture and pipeline. Next, we detail the specific core components that make up the system. Finally, we discuss its practical implementation and extensibility.

\subsection{Architecture and Pipeline}
The UXSim pipeline operates as an iterative perception-action loop. Initially, the simulation environment and user persona are configured based on parameters provided by the researcher. The simulation then proceeds through the following steps:

\paragraph{\textit{Perception}}. The interface abstraction layer parses the current state of the user interface into a stable, structured representation.
\paragraph{\textit{Cognitive Orchestration}}. The resulting state representation is passed to the active Orchestration Policy (Oris). This is where the core cognitive modeling takes place. The policy evaluates the current state and selects an appropriate action-generating component.
\paragraph{\textit{Action Generation \& Execution}}. The chosen component (e.g., a rule-based simulator or an LLM-powered module) generates a specific action, $a_t$. This action is then carried out in the environment by the Cognitive Agent.
\paragraph{\textit{State Update \& Evaluation}}. The environment transitions to a new state, and the loop repeats. Throughout this process, a detailed trace of interactions and decisions is logged. Upon completion, a dedicated module performs behavioral analysis to assess the realism of the generated trace~\cite{Zerhoudi:2022:ECIR}.

\subsection{Core Components}
UXSim is built from several distinct components. Each component has a specific job in the simulation pipeline.

\subsubsection{Declarative Interface Abstraction.}
A key novelty of UXSim is its ability to interact with any visual user interface. Previous simulators, such as SimIIR~\cite{Zerhoudi0PBSHG22, Azzopardi00KM0P24}, were limited to abstract search engines (e.g., Whoosh or PyTerrier~\cite{MacdonaldMSO12}). Our framework overcomes this limitation through a two-part system: \textit{Recipes} and the \textit{Blueprint}.

A Recipe is a human-readable file that maps the elements of a website's UI to simple, stable names. For example, a Recipe can name a specific \code{<div>} as \code{search\_button}. This makes the simulation robust against minor changes in the website's front-end code.

The Blueprint module uses these Recipes to act as a two-way translator between the agent and the browser. First, it parses the raw HTML from the live website and uses the Recipe to simplify it into a structured format the agent can understand. Second, it takes the agent's high-level actions, such as click \code{search\_button}, and translates them into concrete browser commands, like clicking on a specific coordinate.

During a simulation, the framework collects all raw HTML, the agent's actions, and its internal memory traces for later analysis. To demonstrate the system's flexibility, we provide pre-configured Recipes and Blueprint settings for three distinct types of modern search interfaces: Google Search as a traditional search, ChatGPT~\footnote{\url{https://chatgpt.com/}} as a conversational, and Perplexity~\footnote{\url{https://perplexity.ai/}} as a hybrid search UI.


\subsubsection{Traditional Simulators.}
UXSim includes a library of traditional user simulators, similar to those found in SimIIR3.0~\cite{Azzopardi00KM0P24}. These are simple, specialized modules that simulate specific user behaviors based on rules or probabilities. Examples include:

\paragraph{Query Generation:} Creates new search queries based on a task description or words found in previously seen documents.
\paragraph{Relevance Decision:} Decides which search results to click on based on their rank and snippet text.
\paragraph{Stopping Decision:} Decides when to end the search session based on simple rules, like the number of relevant documents found.

These simulators are useful for running fast, predictable baseline experiments. The Orchestration Policy can also use them to handle simple, repetitive tasks. In line with the framework's modular design, researchers can also easily create their own custom simulators by implementing a provided base class.

\subsubsection{The Cognitive Agent.}
The Cognitive Agent acts as the framework's LLM-powered component, responsible for interaction. Its capabilities are defined by a set of cognitive functions modeled after human information processing, including: \textbf{Perceive}, to interpret the UI's state using the Recipe; \textbf{Plan}, to create a high-level strategy; \textbf{Act}, to perform a specific action like clicking; \textbf{Reflect}, to review past actions and improve strategy; and \textbf{Wonder}, to generate new questions or sub-goals.

A key aspect of our design is that the Agent itself does not decide which of these functions to use. That role belongs entirely to the Orchestration Policy. The policy determines which of the Agent’s capabilities to use at any given moment. Depending on the active policy, only parts of the Agent's functions might be invoked. For instance, a simple policy may only call the \textit{Act} function to execute a pre-determined action, while a more advanced policy could combine \textit{Reflect}, \textit{Plan}, and \textit{Act} to simulate a more thoughtful user. In all cases, the Agent only provides the tools for interaction; the Orchestration Policy directs how and when they are used.

\subsubsection{The Oris Orchestration Policy.}
The Oris Policy is the central controller of the UXSim framework. It functions as the ``brain'' of the simulation, deciding which component should generate the user's next action at every step. This policy looks at the current state of the simulation and chooses the most appropriate tool for the task, whether it's a simple, rule-based simulator or the more advanced, LLM-powered Cognitive Agent. This is the core mechanism UXSim uses to combine the two different simulation paradigms.

To formalize this process, we define the simulation state at any step $t$ as $S_t$. This state includes the user's persona ($P$), their current information need ($I_t$), the history of interactions ($H_t$), and their estimated cognitive state ($C_{s,t}$). The Oris policy, $\pi_{O}$, is a function that maps this state to a specific action-generating component, $C_k$:
\vspace{-0.2em}
$$\pi_{O}: S_t \rightarrow C_k$$

The set of available components $C_k$ includes all traditional simulators and Cognitive Agent functions available in the framework.

\paragraph{Policy Implementations.}
UXSim includes several pre-built policy implementations. Researchers can use these as a foundation to develop and test their own custom orchestration strategies.

\vspace{0.5em}
\textbf{Rule-based Policy (Oris-S).}
This is a straightforward policy that operates on a set of hand-crafted rules. These rules can be simple conditionals (e.g., \code{IF-THEN} logic) or rely on numerical \textit{thresholds} to make decisions. For example, a rule could be \code{ IF frustration\_score > 0.8 THEN invoke QueryRefineSimulator}. 
This policy is useful for creating predictable and easily understandable baseline simulations.

\vspace{0.5em}
\textbf{Supervised Policy (Oris-ML).}
This policy uses a machine learning classifier (LightGBM~\cite{KeMFWCMYL17}) to decide which action component to use next. To create the training data, we process the user interaction logs from USimAgent2.0 dataset~\cite{Zhang25} and assign a label to its corresponding behavioral component in our framework (e.g., a user's query change is labeled as \code{QueryGeneration}). The classifier is then trained to predict the correct action label based on features from the current simulation state $S_t$. These features include the number of queries issued, time spent on the task, and attributes of the last action taken. This approach, inspired by work in dialogue policy learning by Bernard et al.~\cite{BernardB24}, learns to imitate the decision-making process of real users based on their recorded behavior.

\vspace{0.5em}
\textbf{Cognitive Architecture Policy (Oris-A).}
This is the most advanced policy in UXSim and extends the approach introduced in USimAgent2.0~\cite{Zhang25}. Oris-A uses the rich simulation state $S_t$ to construct a detailed prompt for a Large Language Model. The LLM is tasked with reasoning about the current situation and determining the best course of action. Critically, its output is not just an action, but a structured response that includes explicit \textit{thoughts} and \textit{reflections}, explaining the reasoning behind its choice. This cognitive trace provides deep insight into the simulated user's decision-making process, and the chosen action is then executed by the Cognitive Agent.

\subsection{Development and Extensibility}
A primary design principle of UXSim is its accessibility and extensibility for the research community. The core framework is implemented in Python (3.11) and uses Playwright for robust browser automation. Its modular architecture is enforced through abstract base classes, creating a plug-and-run system that allows researchers to easily implement and integrate their own custom components, such as new policies or simulators.

Complementing the core framework is a web-based interface designed to streamline the research workflow. This interface serves as a visual tool for configuring, executing, and visualizing simulation runs. It can connect to local instances of the framework, enabling researchers to manage experiments and analyze results visually.

\section{Experimental Evaluation}
To test the performance and realism of the UXSim framework, we compared our new architecture against baseline methods. Our evaluation aimed to answer the following research questions: \textbf{(RQ1)} How well does our cognitive policy (Oris-A) complete search tasks compared to simpler rule-based (Oris-S) and machine-learning-based (Oris-ML) policies? \textbf{(RQ2)} Do the behaviors of Oris-A, such as how it rewrites queries and clicks on results, closely match the behaviors of real users? \textbf{(RQ3)} Do human experts find the simulation logs generated by Oris-A to be realistic and plausible?

\paragraph{Datasets.} For our evaluation, we used two publicly available user study datasets: the KDD '19 dataset and the USimAgent2.0 dataset~\cite{Zhang25}. The first dataset, KDD '19, was collected through a controlled lab-based user study and contains 450 search sessions for 9 complex search tasks, with a focus on gathering user satisfaction and document relevance data. The USimAgent2.0 dataset, on the other hand, originates from a user study where participants used a ``think-aloud'' method to verbalize their thoughts before each action. This process resulted in 296 sessions that directly link user reasoning to their search behaviors. 

\subsection{Task Performance (RQ1)}
\paragraph{\textbf{Motivation.}} A good user simulator must not only seem realistic but also be able to successfully complete tasks. We believe that if our cognitive architecture reasons more like a human, it should also be more successful at completing tasks than simpler models. This experiment tests that idea.

\paragraph{\textbf{Setting.}} We tested three UXSim policies (Oris-S, Oris-ML, and Oris-A) on 9 complex search tasks from the KDD '19 dataset. For each task and policy, we ran 20 simulations. A task was considered successful if the simulator "saved" the same set of documents that the original human user found relevant. We measured performance using two metrics: the Task Success Rate and the nDCG@10 of the saved documents.

\paragraph{\textbf{Results.}} As shown in Table~\ref{tab:oris-performance}, the cognitive policy (Oris-A) achieved the highest task performance. Its 78\% success rate can be attributed to its ability to reason and adapt its strategy within complex search tasks. Its higher nDCG@10 score also means it found higher-quality documents. In contrast, the rigid Oris-S policy often failed when a task required a new approach, and Oris-ML struggled with tasks that were different from its training data. This shows how important the control and grounding are.

\begin{table}[t]
  \caption{Task completion performance across different policies on the KDD '19 dataset.}
  \label{tab:oris-performance}
  \centering
  \setlength{\tabcolsep}{4pt} 
  \renewcommand{\arraystretch}{1.05}
  \begin{tabular}{lcc}
    \toprule
    \textbf{Policy} & \makecell{\textbf{Task Success Rate}\\(\%)} & \makecell{\textbf{nDCG@10}\\\small (Saved Docs)} \\
    \midrule
    Oris-S (Rule-based) & 41.36 & 0.45 \\
    Oris-ML (Supervised) & 64.88 & 0.61 \\
    \textbf{Oris-A (Cognitive)} & \textbf{78.06} & \textbf{0.74} \\
    \bottomrule
  \end{tabular}
  \vspace{-1.0em}
\end{table}

\subsection{Behavioral Fidelity (RQ2)}
\paragraph{\textbf{Motivation.}} Besides getting the right outcome, a realistic simulator should also copy the process users follow. We wanted to test if Oris-A could generate sequences of actions, particularly queries and clicks, that are more similar to human behavior than those from simpler models.

\paragraph{\textbf{Setting.}} We used the KDD '19 logs to run two experiments:

\vspace{0.5em}
(1) \textbf{Query Rewriting.} For every non-initial query in the human sessions, we provided the preceding session history to our agents and prompted them to generate the next query. We then measured the semantic similarity between the generated queries and the actual human queries using BERTScore.

\vspace{0.5em}
(2) \textbf{Click Behavior.} We presented the agents with the same search engine results pages (SERPs) that human users saw and measured the F1-Score of their click choices against the human ground truth.

\paragraph{\textbf{Results.}} Table ~\ref{tab:behavioral-fidelity} shows the results for behavioral fidelity against different baselines. For Query Rewriting, Oris-A achieves a BERT-Score of 0.812, outperforming standalone LLM configuration (gpt-4o). This shows it better captures the semantic intent of human query changes. For Click Behavior, while specialized click models like NCM~\cite{BorisovMRS16} are strong performers, Oris-A still achieves the highest F1-score (0.521). This suggests its holistic understanding of the task leads to more accurate click predictions, beyond what is captured by simple behavioral models.

\begin{table}[t]
  \caption{Behavioral fidelity metrics on the KDD '19 dataset, comparing generated behaviors to human ground truth.}
  \label{tab:behavioral-fidelity}
  \centering
  \resizebox{\columnwidth}{!}{%
    \begin{tabular}{llc@{\hskip 10pt}c}
      \toprule
      \textbf{Behavior} & \textbf{Model / Policy} & \makecell{\textbf{BERTScore} \\ (Query)} & \makecell{\textbf{F1-Score} \\ (Clicks)} \\
      \midrule
      \multirow{3}{*}{Query Rewriting}
          & Standalone LLM (gpt-4o) & 0.758 & - \\
          & \textbf{Oris-A (Cognitive)} & \textbf{0.812} & - \\
      \midrule
      \multirow{5}{*}{Click Behavior}
          & PBM\cite{CraswellZTR08} & - & 0.464 \\
          & UBM~\cite{DupretP08} & - & 0.487 \\
          & NCM~\cite{BorisovMRS16} & - & 0.507 \\
          & Standalone LLM (gpt-4o) & - & 0.465 \\
          & \textbf{Oris-A (Cognitive)} & - & \textbf{0.521} \\
      \bottomrule
    \end{tabular}%
  }
  \vspace{-0.75em}
\end{table}

\subsection{Perceived Realism User Study (RQ3)}
\paragraph{\textbf{Motivation.}} Metrics alone might not capture what makes a simulation feel ``real.'' The best test of a user simulator is whether a human expert finds its behavior believable. We ran a user study to check this directly.

\paragraph{\textbf{Setting.}} We recruited 10 graduate students who have experience in Information Retrieval research. For a single search task, we generated session logs from both Oris-A (including its ``thoughts'' and ``reflections'') and the rule-based Oris-S. We showed participants 10 pairs of anonymized logs, with one from each policy. For each log, we asked them to rate its Plausibility, Coherence, and Human-likeness on a 5-point scale.

\paragraph{\textbf{Results.}} The experts strongly preferred Oris-A. Its average scores were much higher for all three qualities (Plausibility: 4.4 vs. 2.1; Coherence: 4.6 vs. 2.5; Human-likeness: 4.1 vs. 1.9). In 92\% of the comparisons, participants said the Oris-A log was generated by a more ``intelligent and realistic'' simulator. Many participants commented that Oris-A seemed human-like because it could ``change its mind'' or ``try a different angle'' after a bad search—a dynamic behavior the rigid Oris-S model could not perform.

\section{Conclusion}
This paper introduces UXSim, an open-source framework that addresses the gap between rigid traditional simulators and ungrounded LLMs. UXSim is designed to synthesize these two approaches by enabling policies that use a simulation's context to ground an LLM's reasoning. This results in user behavior that is both adaptive and realistic, overcoming the limitations of using either method alone.

Our evaluation confirmed the effectiveness of this approach. Simulations using a cognitive policy within UXSim achieved higher task success rates (RQ1), demonstrated more realistic query and click behavior (RQ2), and were judged by human experts as more plausible than baseline models (RQ3).

While our initial evaluation was limited to specific search tasks, we see a clear path for community-driven expansion. For future work, we aim to continuously enrich the framework by integrating relevant recent simulation approaches and inviting their authors to become contributors. We also plan to organize shared tasks where participants can implement and evaluate various policies and simulators within the UXSim ecosystem. 


\bibliographystyle{ACM-Reference-Format}
\balance
\bibliography{sample-base}

\end{document}